\documentclass[11pt]{article}
\pdfoutput=1
\usepackage[T1]{fontenc}
\usepackage{epsfig,amsfonts,amsthm}
\usepackage{amsmath,amssymb}
\usepackage[normalem]{ulem}
\usepackage{color}
\usepackage{array}
\usepackage{multirow}
\usepackage{hyperref}
\newcommand{\be}{\begin{equation}}
\newcommand{\ee}{\end{equation}}
\newcommand{\bea}{\begin{eqnarray}}
\newcommand{\eea}{\end{eqnarray}}

\def\lsim{\mathrel{\rlap{\lower4pt\hbox{\hskip1pt$\sim$}}
    \raise1pt\hbox{$<$}}}         
\def\gsim{\mathrel{\rlap{\lower4pt\hbox{\hskip1pt$\sim$}}
    \raise1pt\hbox{$>$}}}         
    
\usepackage{amsmath}
\usepackage{amsfonts}
\usepackage{amssymb}
\usepackage{subfig}
\usepackage{wrapfig}
\usepackage{graphicx}

\def\beq{\begin{equation}}
\def\eeq{\end{equation}}
\def\bea{\begin{eqnarray}}
\def\eea{\end{eqnarray}}

\def\<{\left\langle}
\def\>{\right\rangle}

\newcommand{\bt}{\begin{tabular}}
\newcommand{\et}{\end{tabular}}

\usepackage{tabu}

\usepackage{slashed}
\usepackage{amssymb} 
\usepackage{float}

\usepackage[justification=centering]{caption}

\usepackage{tikz}
\usetikzlibrary{decorations.pathmorphing,decorations.markings}

\usepackage{graphicx}

\tikzset{
photon/.style={decorate, decoration={snake,amplitude=2pt, segment length=5pt}, draw=black},
particle/.style={draw=black, postaction={decorate}, decoration={markings,mark=at position .5 with {\arrow[draw=black]{>}}}},
antiparticle/.style={draw=black, postaction={decorate}, decoration={markings,mark=at position .5 with {\arrow[draw=black]{>}}}},
gluon/.style={decorate, draw=black, decoration={coil,amplitude=4pt, segment length=5pt}}
goldstone/.style={draw=green,postaction={decorate},decoration={markings,mark=at position .5 with {\arrow[draw=blue]{>}}}}
}

\definecolor{grey}{cmyk}{0,0,0,0.75}
\definecolor{tangerine}{cmyk}{0,0.5,1,0}
\definecolor{darkgreen}{cmyk}{1,0,1,0.23}
\definecolor{Red}{rgb}{1,0,0}
\definecolor{Blue}{rgb}{0,0,1}
\definecolor{Green}{cmyk}{1,0,1,0.23}

\newcommand{\green}[1]{#1}
\newcommand{\red}[1]{#1}

 \usepackage[
top    = 3.5cm,
bottom = 3.5cm,
left   = 3.5cm,
right  = 3.5cm]{geometry}

\allowdisplaybreaks[2]
\begin{document}
\bibliographystyle{OurBibTeX}

\title{\hfill ~\\[-30mm]
                  \textbf{Natural neutrino sector in a 331-model with Froggatt-Nielsen mechanism
                }        }
\author{\\[-5mm]
Katri Huitu\footnote{E-mail: {\tt Katri.Huitu@helsinki.fi}},\ 
Niko Koivunen\footnote{E-mail: {\tt Niko.Koivunen@kbfi.ee}},\  
Timo \green{J.} K{\"a}rkk{\"a}inen\footnote{E-mail: {\tt karkkainen@caesar.elte.hu}}\  
 \\
  \emph{\small Department of Physics and Helsinki Institute of Physics,}\\
  \emph{\small Gustaf H{\"a}llstr{\"o}min katu 2, \green{FI-00014} University of Helsinki, Finland}\\[4mm]}
\maketitle

\vspace*{-0.250truecm}
\begin{abstract}
\noindent
{
The extensions of the Standard Model  based on the $SU(3)_c\times SU(3)_L\times U(1)_X$ gauge group (331-models) have been advocated to explain the number of fermion families in nature. It has been recently shown that the Froggatt-Nielsen mechanism, a popular way to explain the mass hierarchy of the charged fermions, can be incorporated into the 331-setting in an economical fashion (FN331). In this work we extend the FN331-model to include three right-handed neutrino singlets. We show that the seesaw mechanism is realized in this model. The scale of the seesaw mechanism is near the $SU(3)_L\times U(1)_X$-breaking scale. The model we present here simultaneously explains the mass hierarchy of all the fermions, including neutrinos, and the number of families.   

} 
\end{abstract}
 
\thispagestyle{empty}
\vfill
\newpage
\setcounter{page}{1}

\section{Introduction}
After the discovery of the Higgs boson at the Large Hadron Collider, the last elementary particle predicted by the  Standard Model (SM) has been confirmed. Today, particle physics has moved on to a new era, where we  attempt to answer the problems plagueing the SM with  economical extensions. The problems include number of generations, nonzero neutrino mass, neutrino mixing and fermion mass hierarchies.

In Nature three generations of quarks and leptons have been observed. Number of neutrino flavours is $2.984 \pm 0.008$ \cite{Tanabashi:2018oca}-\cite{Akrawy:1990zy}, which is a statistical fit to SM using LEP data. This is a strong indication for exactly three generations of matter, which, however, is not imposed by SM itself. We know from neutrino oscillation experiments that at least two of the three SM neutrinos are massive, with masses less than $0.1$ eV and the sum of their masses is less than  $0.12$ eV from cosmological constraints by the PLANCK experiment. Neutrino masses are not included in the Standard Model, and they are six orders of magnitude lighter than the next lightest massive particle, electron, and twelve orders of magnitudes lighter than the heaviest particle, top quark. This huge range of different masses gives birth to the flavour problem.

Extensions of the Standard Model based on the $SU(3)_c\times SU(3)_L\times U(1)_X$ gauge group (331-models) have been proposed in the  literature to explain the number of fermion families in Nature. In the traditional 331-models  \cite{Pisano:1991ee}-\cite{Dong:2013ioa}  the gauge anomalies cancel only if the number of fermion familes is three. 
The  $SU(3)_c\times SU(3)_L\times U(1)_X$ gauge group contains one additional diagonal generator compared to the SM-gauge group  $SU(3)_c\times SU(2)_L\times U(1)_Y$. This means that the electric charge can be defined in multiple different ways in the 331-models:
\be
Q=T_3+\beta T_8+X,
\ee
where the parameter $\beta$ defines the particle content of the model. The models with $\beta=\pm \sqrt{3}$ \cite{Pisano:1991ee}-\cite{Nguyen:1998ui} and $\beta= \pm 1/\sqrt{3}$ \cite{Georgi:1978bv}-\cite{Dong:2013ioa}\footnote{The model presented in \cite{Georgi:1978bv} does not exhibit the cancellation of gauge anomalies and does not explain the number of fermion families.}  are extensively  studied in the literature. The models with $\beta=\pm \sqrt{3}$ contain particles  with exotic electric charges such as doubly charged \red{scalars and} gauge bosons. They also contain a very  large scalar sector, composed of three $SU(3)_L$-triplets and an $SU(3)_L$-sextet.  The models based on the  $\beta= \pm 1/\sqrt{3}$ on the other hand have simpler scalar sector, composed from only three $SU(3)_L$-triplets. The models based on  $\beta= \pm 1/\sqrt{3}$ do not contain particles with exotic electric charges.   Also the models with $\beta=0$ have been studied \cite{Hue:2015mna}.

Even though the 331-models can shed light on  the number of fermion familes, the fermion mass hierarchy is left unexplained in the traditional models.
Recently it was  shown that  the Froggatt-Nielsen mechanism \cite{Froggatt:1978nt} can be incorporated into the 331-models with $\beta=\pm 1\sqrt{3}$ without extending the scalar sector \cite{Huitu:2017ukq,Huitu:2019kbm}. The Froggatt-Nielsen mechanism (FN) is a well established method to explain the mass hierarchy of the fermions, and   a 331-model with incorporated FN-mechanism (FN331) can therefore   simultaneously explain both the number of fermion families and the mass hierarchy of the charged fermions. The neutrino masses and mixings are not naturally explained in FN331-model, however. 
The neutrino mass matrix is antisymmetric in the FN331-model and therefore one of the neutrinos is  massless and the two other mass degenerate at tree-level. Loop corrections are needed to lift the one eigenvalue from zero and to break the degeneracy of the other two \cite{Valle:1983dk}. This neutrino sector is identical to the one presented in \cite{Singer:1980sw}-\cite{Dong:2013ioa}.  

Our aim is to extend the neutrino sector to make it natural and explain the neutrino masses and mixings without fine-tuning at tree-level.  We propose an extension of the FN331-model where we add three right-handed neutrino singlets  to the model. This allows the tree-level masses for all the neutrinos and  implementation of seesaw mechanism \cite{Minkowski:1977sc}-\cite{Glashow:1979nm} for the neutrino sector. Here the seesaw is combined with the FN-mechanism, which allows the seesaw scale to be low thanks to suppression in the neutrino Yukawa couplings due to the  FN-mechanism. 
The seesaw scale is essentially the same as the $SU(3)_L\times U(1)_X$-breaking scale. We study as low $SU(3)_L\times U(1)_X$-breaking scale as possible.  The suppression of \emph{flavour changing neutral currents} (FCNC) allows for $SU(3)_L\times U(1)_X$-breaking scale as low as 5 TeV as shown in \cite{Huitu:2019kbm}. The collider bounds on the $Z'$ mass however suggest that the   $SU(3)_L\times U(1)_X$-breaking scale has to be at least around 7 TeV \cite{Salazar:2015gxa}.  We aim to generate singlet neutrinos at medium-energy scale and sterile neutrinos at TeV scale, utilizing seesaw mechanism. By medium-energy we refer to a mass scale between the active neutrinos and mass of electron. We will present the lowest possible SU(3)$_L \times $U(1)$_X$ breaking scale consistent with experimental data, which turns out to be approximately 7 TeV.

The hierachical structure of the neutrino Yukawa couplings is determined by the FN-mechanism. The hierarchy of the neutrino Yukawa couplings can be arranged so that the neutrino mixings are generated without fine-tuning. 
All the PMNS-matrix elements are experimentally known to be order-one-numbers. This kind of pattern of mixing can be achieved in FN-setting by  assigning  all the FN-charges of the left-handed leptons to be equal. We show that the correct sub-eV neutrino masses, mass square differences and mixing angles  are produced within experimental limits.

The paper is structured as follows. We present the particle content of the model in  Section \ref{particle content}. In Section \ref{Yukawa sector} we  review the Froggatt-Nielsen mechanism in the 331-setting, the FN331-model. In Sections \ref{Charged lepton Yukawa couplings and masses} to \ref{Neutrino coupling to charged gauge bosons and PMNS-matrix} we study the lepton mass matrices and mixings and finally in the Section \ref{constraints and numerical example} we present numerical example for the neutrino masses and mixings.

\section{Particle content}\label{particle content}
We propose a model where the gauge group of the Standard Model  is extended to $SU(3)_C\times SU(3)_L\times U(1)_X$. We define the electric charge as\footnote{The choice $\beta=+\frac{1}{\sqrt{3}}$ would result in essentially a same model.}:

\begin{equation}
Q=T_3-\frac{1}{\sqrt{3}}T_8+X,
\end{equation}
where the $T_3$ and $T_8$ are the diagonal $SU(3)_L$ generators. 
We also introduce global $U(1)_{FN}$-symmetry, under which fermions and some of the scalars are charged. 

\subsection{Fermion representations}\label{fermion representations}
 Let us now write down the fermion representations. The left-handed leptons are assigned to  $SU(3)_{L}$ -triplets and the right-handed leptons are assigned to $SU(3)_{L}$-singlets:

\begin{eqnarray}
&&L_{L,i}=\left(\begin{array}{c}
\nu_{i}\\
e_{i}\\
\nu'_{i}
\end{array}
\right)_{L}\sim (1,3,-\frac{1}{3}), \quad i=1,2,3,\\
&&  e_{R,i}\sim (1,1,-1), \quad  N_{R,i}\sim (1,1,0).
\end{eqnarray}
 The numbers in the parantheses label the transformation propeties under the gauge group  $SU(3)_{c}\times SU(3)_{L}\times U(1)_{X}$. The $\nu'_{L,i}$ and $N_{R,i}$ are new leptons with electric charges $0$. The three right-handed neutrinos $N_{R,i}$ are not present in the model studied in  \cite{Huitu:2017ukq,Huitu:2019kbm} and they allow the tree-level masses  for all the neutrinos.

The cancellation of anomalies requires the number of fermion triplets to be the same as antitriplets. This is achieved by assigning two quark families to $SU(3)_{L}$ antitriplets and one family to a triplet. We choose to assign first quark generations into triplet and the second and the third  into an antitriplet:

\begin{eqnarray}
&&Q_{L,1}=\left(\begin{array}{c}
u_{1}\\
d_{1}\\
U
\end{array}
\right)_{L}\sim (3,3,\frac{1}{3}),\\
&&Q_{L,2}=\left(\begin{array}{c}
d_{2}\\
-u_{2}\\
D_{1}
\end{array}
\right)_{L},\quad 
Q_{L,3}=\left(\begin{array}{c}
d_{3}\\
-u_{3}\\
D_{2}
\end{array}
\right)_{L}\sim (3,3^{\ast},0),\\
&&  u_{R,i}\sim (3,1,\frac{2}{3}), \quad  U_{R}\sim (3,1,\frac{2}{3}),\\
&&  d_{R,i}\sim (3,1,-\frac{1}{3}), \quad  D_{R,1}\sim (3,1,-\frac{1}{3}), \quad  D_{R,2}\sim (3,1,-\frac{1}{3}),\quad i=1,2,3.
\end{eqnarray}
 
We have introduced new quarks $D_{1}$ and  $D_{2}$ with electric charge $-1/3$ and $U$ with electric  charge $2/3$, which will mix with the SM quarks.  All the fermions are also charged under the  global Froggatt-Nielsen $U(1)_{FN}$ symmetry. This will be discussed in detail in the dedicated Section \ref{Yukawa sector}.

 When we take into account the colour, there are six fermion triplets and six antriplets, ensuring the cancellation of pure $SU(3)_L$-anomaly. All the gauge anomalies will cancel with this particle content. The anomaly cancellation forces one quark generation to be placed in a different representation than the other two. The unequal treatment of quark generations inevitably leads to scalar mediated  FCNCs at tree-level, which is a feature plaguing the traditional 331-models \cite{Pisano:1991ee}-\cite{Dong:2013ioa}. It was however recently shown that tree-level scalar mediated FCNCs of quarks are suppressed in the FN331-model \cite{Huitu:2019kbm}. This is in contrast to the traditional 331-models, which offer no natural suppression mechanism for the tree-level  scalar mediated FCNCs.



\subsection{Scalar sector}\label{scalar sector}
The 331-models with $\beta=-\frac{1}{\sqrt{3}}$ contain only two types of scalar triplets with neutral entries: $X=2/3$ and $X=-1/3$. One must include at least  two $X=-1/3$ triplets with $X=-1/3$  and one  triplet with $X=2/3$ in order to generate the masses for all the charged fermions at tree level. We choose to have this minimal scalar sector:

\begin{eqnarray}
&&\eta=\left(\begin{array}{c}
\eta^{+}\\
\eta^{0}\\
{\eta'}^{+}
\end{array}
\right)\sim (1,3,\frac{2}{3}),\quad  
\rho=\left(\begin{array}{c}
\rho^{0}\\
\rho^{-}\\
{\rho'}^{0}
\end{array}
\right)\sim (1,3,-\frac{1}{3}),\label{scalar triplets}\\
&&\chi=\left(\begin{array}{c}
\chi^{0}\\
\chi^{-}\\
{\chi'}^{0}
\end{array}
\right)\sim (1,3,-\frac{1}{3}).\nonumber
\end{eqnarray}
 
All  the neutral fields can in general develop a nonzero vacuum expectation value (VEV).  The minima are related to each other  by $SU(3)_L$ rotation. We choose to rotate  one of the $X=-1/3$ triplet VEVs so that the upper component VEV goes to zero. This rotation will leave the rest of the VEVs general. So we  have  vacuum structure:  

\begin{eqnarray}\label{vacuum}
&&\langle\eta\rangle=\frac{1}{\sqrt{2}}\left(\begin{array}{c}
0\\
v'\\
0
\end{array}
\right), \quad
\langle\rho\rangle=\frac{1}{\sqrt{2}}\left(\begin{array}{c}
v_1\\
0\\
v_2
\end{array}
\right),
\quad
\langle\chi\rangle=\frac{1}{\sqrt{2}}\left(\begin{array}{c}
0\\
0\\
u
\end{array}
\right).\label{vacuum}
\end{eqnarray}  
The VEVs $v_2$ and $u$ break the $SU(3)_L$-symmetry, and we assume them to be: $v_2,u\gtrsim \mathcal{O}(\textrm{TeV})$. The VEVs $v'$ and $v_1$ break the $SU(2)_L$ symmetry and we take them to be of the order of the electroweak scale.

The scalar triplets in Eq. (\ref{scalar triplets}) are charged under the global symmery $U(1)_{FN}$ with the charge assignment presented in the Table \ref{FN charges of scalars}. Note that since the scalar triplets $\rho$ and $\chi$ are in the same representation, the combination $\rho^\dagger\chi$ is gauge invariant. Also according to Eq. (\ref{vacuum}) and the Table \ref{FN charges of scalars}, the gauge invariant combination  $\rho^\dagger\chi$ carries a non-zero FN-charge and has a non-zero VEV. Therefore the  $\rho^\dagger\chi$-combination can play the role of the  flavon in the Froggatt-Nielsen mechanism, as was demonstrated in \cite{Huitu:2017ukq,Huitu:2019kbm}. The Froggatt-Nielsen mechanism can thus be implemented without introducing new scalar degrees of freedom into the model.

\begin{table}[h!]
\begin{center}
\begin{tabular}{|c|c|c|c|}
\hline
Particle   &  $\eta$ & $\rho$ & $\chi$ \\
\hline 
FN-charge & $-1$ & $1$ & $0$ \\
\hline
\end{tabular}
\end{center}
\caption{The FN $U(1)$ charges of the scalars.}
\label{FN charges of scalars}
\end{table}

The scalar potential is greatly simplified due to inclusion of global $U(1)_{FN}$-symmetry. 
 The most general $U(1)_{FN}$-symmetric scalar potential is,
\begin{eqnarray}
&& V_{\textrm{FN}}=\mu^2_1 \eta^{\dagger}\eta+\mu_2^2 \rho^{\dagger}\rho+\mu_3^{2}\chi^{\dagger}\chi
+\lambda_1 (\eta^{\dagger}\eta)^2+\lambda_2 (\rho^{\dagger}\rho)^2+\lambda_{3}(\chi^{\dagger}\chi)^2\label{FN symmetric potential}\\
&& +\lambda_{12} (\eta^{\dagger}\eta)(\rho^{\dagger}\rho)+\lambda_{13} (\eta^{\dagger}\eta)(\chi^{\dagger}\chi)+\lambda_{23}(\rho^{\dagger}\rho)(\chi^{\dagger}\chi)\nonumber\\
&& +\widetilde{\lambda}_{12} (\eta^{\dagger}\rho)(\rho^{\dagger}\eta)+\widetilde{\lambda}_{13} (\eta^{\dagger}\chi)(\chi^{\dagger}\eta)+\widetilde{\lambda}_{23}(\rho^{\dagger}\chi)(\chi^{\dagger}\rho)\nonumber\\
&&+\sqrt{2}f(\epsilon_{ijk}\eta^{i}\rho^{j}\chi^{k}+h.c.).\nonumber
\end{eqnarray}
However the global $U(1)_{FN}$-symmetry is spontaneously broken by the scalar field VEVs.  This leaves one Golstone boson to the physical spectrum. In order to give it  a mass we add the following soft FN-breaking term to the potential:
\begin{equation}
V_{\textrm{soft}}=b(\rho^{\dagger}\chi)+h.c.
 \end{equation}

All the complex phases in the scalar potential can be absorbed into the fields and therefore all the parameters in the scalar potential are real. 
The real and imaginary parts of the scalars will therefore not  mix. We choose the parameter $f$ to be comparable to the $SU(3)_L\times U(1)_X$-breaking scale. The soft-breaking term is also chosen to be large, $b\sim -(v_\text{heavy})^2$, in order to decouple the pseudo-Goldstone boson in the low energies.

  The scalar sector has five CP-even, five CP-odd  and four charged  scalars. One CP-even, three CP-odd and two charged scalars are massless would-be-Goldstone bosons that are absorbed by the gauge bosons of the model, namely the $Z$, $W^\pm$ of the SM, new heavy charged gauge boson $V^\pm$, new heavy neutral gauge boson $Z'$ and a non-hermitian heavy neutral gauge boson $X^0$. Thus there are four CP-even, two CP-odd and two charged scalars left in the physical spectrum. All the physical scalars, except the 125 GeV scalar, have their masses around the $SU(3)_L\times U(1)_X$-breaking scale and they are  very heavy. The details of the scalar mass matrices are provided in the Appendix \ref{scalar masses}.


\subsection{Gauge sector}
As previously mentioned the gauge sector of 331-model is enlarged compared to the SM. The 331-models will contain five additional gauge bosons compared to the SM. 
The covariant derivative for $SU(3)_L$ triplet is:
\begin{eqnarray}
&&D_{\mu}=\partial_{\mu}-ig_3 \sum_{a=1}^8 T_{a}W_{a\mu}-ig_{x}XB_{\mu},\nonumber
\end{eqnarray}
where  $g_3$ and $g_x$ are the $SU(3)_L$ and $U(1)_X$ gauge couplings respectively. The $T_a=\lambda_a/2$ are the $SU(3)_L$ generators, where $\lambda_a$ are the Gell-Mann matrices.  The $SU(3)_L$ gauge bosons are, 
\begin{eqnarray}
 &&\sum_{a=1}^8 T_a W_{a\mu}=\frac{1}{\sqrt{2}}\left(\begin{array}{ccc}
\frac{1}{\sqrt{2}}W_{3\mu}+\frac{1}{\sqrt{6}}W_{8\mu} &
 {W'}_{\mu}^{+} &  
{X'}^{0}_{\mu}\\
 {W'}^{-}_{\mu} &
-\frac{1}{\sqrt{2}}W_{3\mu}+\frac{1}{\sqrt{6}}W_{8\mu}& 
 {V'}^{-}_{\mu}\\
 X'^{0\ast}_{\mu} & 
 {V'}^{+}_{\mu}&
-\frac{2}{\sqrt{6}}W_{8\mu}
\end{array}\right),\nonumber
\end{eqnarray} 
where we have denoted,
\begin{eqnarray}
{W'}^{\pm}_\mu=\frac{1}{\sqrt{2}}(W_{1\mu}\mp iW_{2\mu}),\nonumber\\
{V'}^{\mp}_\mu=\frac{1}{\sqrt{2}}(W_{6\mu}\mp iW_{7\mu}),\nonumber\\
{X'}^{0}_\mu=\frac{1}{\sqrt{2}}(W_{4\mu}-iW_{5\mu}).\nonumber
\end{eqnarray}
The fields  $W_{3\mu}$, $W_{8\mu}$, $B_\mu$ and $W_{4\mu}$ will form neutral mass eigenstates: photon, $Z$-boson and new heavy gauge bosons $Z'$ and $\widetilde{W}_{4\mu}$. The field $W_{5\mu}$ does not mix with the other neutral gauge bosons and is a mass eigenstate, with same mass as $\widetilde{W}_{4\mu}$. These fields are identified as a \emph{physical neutral non-hermitian gauge boson} $X^0_\mu\equiv\frac{1}{\sqrt{2}}(\widetilde{W}_{4\mu}-iW_{5\mu})$. Details of the neutral gauge boson masses are given in the Appendix \ref{neutral gauge boson masses appendix}.
 The off-diagonal gauge bosons ${W'}^{\pm}_\mu$ and ${V'}^{\pm}_\mu$ will form the SM gauge bosons $W^{\pm}_\mu$ and the heavy new gauge bosons $V^{\pm}_\mu$.

\subsubsection{Charged gauge bosons}\label{charged gauge bosons}
The mass term for the charged gauge bosons is given by,
\be
\mathcal{L}\supset Y^T M^2_{charged} Y,
\ee
where $Y^T=({W'}_{\mu}^{+},{V'}_{\mu}^{+})$ and, 
\begin{equation}
M^2_{charged}=\left(\begin{array}{cc}
\frac{g_3^{2}}{4}({v'}^2+v_1^2)& \frac{g_3^{2}v_1 v_2}{4}\\
\frac{g_3^{2}v_1 v_2}{4} & \frac{g_3^{2}}{4}({v'}^2+v_2^2+u^2)
\end{array}\right),
\end{equation}
is the charged gauge boson mass matrix.
The eigenvalues of the matrix are,
\begin{eqnarray}
&&m_{W^{\pm}}^2
=  \frac{g_3^2}{4}({v'}^2+\frac{v_1^2 u^2}{v_2^2+u^2})+\mathcal{O}\left(\frac{v^2_{\textrm{light}}}{v^2_{\textrm{heavy}}}\right),\label{charged gauge boson masses}\\
&&m_{V^{\pm}}^2
= \frac{g_3^2}{4}(v_2^2+u^2)+\mathcal{O}\left(\frac{v^2_{\textrm{light}}}{v^2_{\textrm{heavy}}}\right),\nonumber
\end{eqnarray}
where $v_\text{heavy}=v_2, u$ and $v_\text{light}=v_1, v'$.
 According to Eq. (\ref{charged gauge boson masses}) the SM Higgs VEV is related to the triplet VEVs through the relation
\begin{equation}\label{relation among vevs}
{v'}^2+\frac{v_1^2 u^2}{v_2^2+u^2}+\mathcal{O}\left(\frac{v^2_{\textrm{light}}}{v^2_{\textrm{heavy}}}\right)=(v_{sm})^2,
\end{equation}
where $v_{sm}=246$ $\textrm{GeV}$.

The mass eigenstates are defined as 
\be
\left\{\begin{array}{c}
W^{+}_{\mu}=\cos\theta ~{W'}^{+}_{\mu}+\sin\theta ~{V'}^{+}_{\mu}\\
V^{+}_{\mu}=-\sin\theta ~{W'}^{+}_{\mu}+\cos\theta ~{V'}^{+}_{\mu}
\end{array}\right. 
,\label{charged gauge boson mass eigenstates}
\ee
where the mixing angle $\theta$ is defined as:
\begin{equation}\label{tan of charged gauge boson mixing angle}
\tan 2\theta=-\frac{2v_1 v_2}{v_2^2+u^2-v_1^2}.
\end{equation}
The mixing angle between $W^\pm_\mu$ and $V^\pm_\mu$ is tiny, $\theta\sim\frac{v_\text{light}}{v_\text{heavy}}$,  due to large difference between the $SU(3)_L\times U(1)_X$ and $SU(2)_L\times U(1)_Y$-breaking VEVs.
The  SM gauge boson $W^{\pm}_{\mu}$ is almost totally  ${W'}^{\pm}_{\mu}$ and  $V^{\pm}_{\mu}$ is mostly  ${V'}^{\pm}_{\mu}$. \green{The experimental bound for the mixing is $|\theta| \lesssim \mathcal{O}(10^{-2})$ \cite{Tanabashi:2018oca}, and has been taken into account in our numerical analysis (Section \ref{constraints and numerical example}).}

\section{The Yukawa sector and the Froggatt-Nielsen mechanism  in the  331-framework}\label{Yukawa sector}
Next we study the Yukawa sector of the model. We are employing the Froggatt-Nielsen mechanism to generate the hierarchical structure of the fermion Yukawa couplings.
The  original Froggatt-Nielsen model extends the Standard Model with a flavour symmetry (FN symmetry), whose symmetry group in the simplest case is global or local $U(1)$ or  a discrete $Z_N$ symmetry. 
The FN-framework introduces a new complex scalar field, the flavon, which is a singlet under standard model gauge group  $SU(3)_C\times SU(2)_L\times U(1)_Y$.
The SM fermions, the  SM Higgs and the flavon  are charged under the FN symmery. The key property  of the FN-symmetry is to forbid the SM Yukawa couplings, save perhaps the top quark. The SM Yukawa couplings are generated as effective couplings instead. 

The FN mechanism can be economically incorporated into a 331-model.
The scalar sector we have introduced in section \ref{scalar sector} contains the neccesary incredients to house Froggatt-Nielsen mechanism, as was demonstrated in \cite{Huitu:2017ukq,Huitu:2019kbm}. The addition of complex scalar field to act as a flavon is thus unneccesary. Here a gauge singlet combination, $\rho^\dagger\chi$, will act as the flavon instead of single complex scalar field. 
The effective flavon, $\rho^\dagger\chi$, obtains a nonzero vacuum expectation value,    $\langle\rho^{\dagger}\chi\rangle=(v_2 u)/2$, as can be seen from the Eq. (\ref{vacuum}).

The effective operator that generates the Yukawa couplings of the charged fermions is\footnote{The effective operator that generates  the neutrino Yukawa couplings is presented later in Eq. (\ref{FN operator for neutrinos}).}:
\begin{equation}\label{usual 331 FN operator}
\mathcal{L}\supset (c_s^{f})_{ij}\left(\frac{\rho^{\dagger}\chi}{\Lambda^2}\right)^{(n^s_f)_{ij}}\bar{\psi}_{L,i}^f S f_{R,j}+h.c.,
\end{equation}
where $(c_s^{f})_{ij}$  is a dimensionless order-one number, $\Lambda$ is the scale of the new physics,  $S$ denotes any of the three scalar triplets $\eta$, $\rho$ or $\chi$. The $\bar{\psi}_{L,i}^f$  and $f_{R,j}$ represent here the fermion triplets, antitriplets and singlets that were introduced in section \ref{fermion representations}. The  power  $(n^s_f)_{ij}$ is positive integer number\footnote{If $(n^s_f)_{ij}$ were negative, we would have to include operator $(c_s^{f})_{ij}\left(\frac{\chi^{\dagger}\rho}{\Lambda^2}\right)^{-(n^s_f)_{ij}}\bar{\psi}_{L,i}^f S f_{R,j}+h.c.$} and   determined by the FN charge conservation (see  Table \ref{FN charges in 331}):
\begin{equation}
(n^s_f)_{ij}=\left[q(\bar{\psi}_{L,i}^f)+q(f_{R,j})+q(S)\right].
\end{equation}

\begin{table}[h!]
\begin{center}
\begin{tabular}{|c|c|c|c|}
\hline
Particle    & $(\psi^f_{L,i})^{c}$ & $f_{R,i}$ & $S$\\
\hline 
FN charge & $q(\bar{\psi}_{L,i}^f)$ & $q(f_{R,j})$ & $q_{S}$ \\
\hline
\end{tabular}
\end{center}
\caption{The FN charges of  fermions and the scalar fields.}
\label{FN charges in 331}
\end{table}

The usual 331-model Yukawa terms are generated as effective couplings when  the scalar triplets $\rho$ and $\chi$ acquire VEVs:
\begin{eqnarray}
&&(c^f_s)_{ij}\left(\frac{(\rho+\langle\rho\rangle)^{\dagger}(\chi+\langle\chi\rangle)}{\Lambda^2}\right)^{(n^s_f)_{ij}}\bar{\psi}_{L,i}^f(S+\langle S\rangle) f_{R,j}+h.c.\label{331 FN}\\
&&=(y^f_s)_{ij}\bar{\psi}_{L,i}^f(S+\langle S\rangle) f_{R,j}
+(n^s_f)_{ij}(y^f_s)_{ij}\left[\frac{{\rho'}^{0\ast}}{v_2}+\frac{{\chi'}^0}{u}
+\frac{v_1 {\chi}^0 }{v_2 u}\right]\sqrt{2}\bar{\psi}_{L,i}^f\langle S\rangle f_{R,j}+h.c.+\cdots,\nonumber
\end{eqnarray} 
where  only the renormalizable contributions are kept. The first term  in  Eq. (\ref{331 FN}) gives  the usual Yukawa terms of the model, as in the original FN model.  
The Yukawa coupling is now defined  as:
\begin{equation}
(y^f_s)_{ij}=(c^f_s)_{ij}\left(\frac{v_2 u}{2\Lambda^2}\right)^{(n^s_f)_{ij}}\equiv 
(c^f_s)_{ij}\epsilon^{(n^s_f)_{ij}}.
\end{equation}
Hierarchical Yukawa couplings are produced by assuming that $\epsilon=(v_2 u) /(2\Lambda^2)<1$.  The FN-charges of the SM fermions determine the power $(n_s^f)_{ij}$ and therefore the amount of suppression each Yukawa coupling obtains.  One can obtain the observed fermion mass hierarchy by assigning larger FN charges to the lighter fermions compared to the heavier ones.  This is  in contrast to the Standard Model where the hierarchy is obtained only by fine-tuning the couplings themselves. 

The second term in Eq. (\ref{331 FN}) is not proportional to the Yukawa matrix  and is therefore flavour violating.
 This flavour violating part is suppressed by the scale of the $SU(3)_L\times U(1)_X$-breaking.  We assume that the $SU(3)_L\times U(1)_X$-breaking scale $v_\mathrm{scale}$ is $\gtrsim 7$ TeV. This is enough to suppress the quark FCNCs as shown in \cite{Huitu:2019kbm}. This also ensures that the $Z'$ mass is $\gtrsim 3$ TeV, satisfying bounds in  \cite{Salazar:2015gxa}.  We will safely  ignore the flavour violating contributions as they are heavily suppressed.

\section{Charged lepton Yukawa couplings and masses}\label{Charged lepton Yukawa couplings and masses}

Charged lepton Yukawa couplings are:

\begin{eqnarray}
&&\mathcal{L}_{\textrm{lepton}}
=y^{e}_{ij}\bar{L}_{L,i}\eta e_{R,j}+l_{ij}y^e_{ij}~\bar{e}_{L,i}e_{R,j}~\left[\frac{v'}{v_2}{\rho'}^{0\ast}+\frac{v'}{u}{\chi'}^{0}+\frac{v'v_1}{v_2 u}{\chi}^{0}\right]
+h.c.,\label{lepton-yukawa}
\end{eqnarray}
where $i,j=1,2,3$. The first term is the traditional 331 Yukawa term for the charged leptons whereas the second term is the additional Yukawa interaction term due to the FN-mechanism. As stated earlier,  this additional term is suppressed by the $v_\text{heavy}$ and we will ignore it in the following. 
The charged lepton  Yukawa matrix is given as follows:
\begin{eqnarray}\label{ce-eqs}
&&y^e_{ij}=c^e_{ij}\epsilon^{q(\bar{L}_{L,i})+q(e_{R,j})+q(\eta)}
\equiv c^e_{ij}\epsilon^{l_{ij}}.
\label{lepton-epsilon}
\end{eqnarray}
We will specify the FN-charges we use later in section \ref{constraints and numerical example}, when we study numerical examples.

The charged leptons acquire masses as the scalar triplet $\eta$ obtains a VEV:
\be
\mathcal{L}\supset y^{e}_{ij}\bar{L}_{L,i}\langle\eta\rangle e_{R,j}+h.c.
=m_{ij}^e\bar{e}_{L,i} e_{R,j}+h.c.,
\ee
where the charged lepton mass matrix is,
\be\label{charged lepton mass matrix}
m_{ij}^e=y^{e}_{ij}\frac{v'}{\sqrt{2}}.
\ee
The charged lepton mass matrix is diagonalized as:
\be\label{charged lepton mass matrix diagonalization convention}
U^e_L m^e U_R^{e\dagger}=m^{e}_{diag}.
\ee
The charged lepton mass matrix proportional to the Yukawa matrix  will be diagonalized simultaneously with the Yukawa coupling. Therefore there will be no flavour changing couplings in the standard Yukawa couplings.  The only flavour violation to the charged leptons is coming from the Froggatt-Nielsen mechanism which is however suppressed.

\section{Neutrino mass matrix}
The neutrino Yukawa couplings originate from effective operators of two types. The first type was already presented in Eq. (\ref{usual 331 FN operator})  and the operator of the second kind is:
\be\label{FN operator for neutrinos}
\mathcal{L}\supset (c_{\eta^\ast}^N)_{ij}\left(\frac{\rho^\dagger\chi}{\Lambda^2}\right)^{(n_{N}^{\eta^\ast})_{ij}}\epsilon_{\alpha\beta\gamma}\bar{L}_{L,i}^{\alpha}(L^{c}_{L,j})^{\beta}(\eta^{\ast})^{\gamma}+h.c.
\ee
The operators in Eqs. (\ref{usual 331 FN operator}) and (\ref{FN operator for neutrinos}) produce the following Yukawa couplings for neutrinos:
\begin{eqnarray}\label{cnu-eqs}
\mathcal{L}_{\textrm{neutrino}}&=&e_{ij}\epsilon_{\alpha\beta\gamma}\bar{L}_{L,i}^{\alpha}(L^{c}_{L,j})^{\beta}(\eta^{\ast})^{\gamma}+{y}^{N}_{ij}\bar{L}_{L,i}\rho N_{R,j}
+{y'}_{ij}^{N}\bar{L}_{L,i}\chi N_{R,j}\nonumber\\
&&+\left\{(n^{\eta^\ast}_N)_{ij}e_{ij}\epsilon_{\alpha\beta\gamma}\bar{L}_{L,i}^{\alpha}(L^{c}_{L,j})^{\beta}\langle\eta^\ast\rangle^\gamma
+n_{ij}y^N_{ij}~\bar{L'}_{L,i}\langle\rho\rangle N'_{R,j}
+n'_{ij}{y'}^N_{ij}~\bar{L'}_{L,i}\langle\chi\rangle N'_{R,j}\right\}\times\nonumber\\
&&\times\sqrt{2}\left[\frac{{\rho'}^{0\ast}}{v_2}+\frac{{\chi'}^{0}}{u}+\frac{v_1}{v_2 u}{\chi}^{0}\right]+h.c.,\label{neutrino standard+FN}
\end{eqnarray}
where $i,j=1,2,3$ and the Yukawa couplings are, 
\begin{eqnarray}
y^N_{ij}&=&c^N_{ij}\epsilon^{q(\bar{L}_{L,i})+q(N_{R,j})+q(\rho)}
=c^N_{ij}\epsilon^{n_{ij}},\nonumber\\
{y'}^N_{ij}&=&{c'}^N_{ij}\epsilon^{q(\bar{L}_{L,i})+q(N_{R,j})+q(\chi)}
={c'}^N_{ij}\epsilon^{n'_{ij}},\label{neutrino Yukawas in terms of epsilon}\\
e_{ij}&=&(c^N_{\eta^\ast})_{ij}
\epsilon^{q(L^c_{L,i})+q(L^c_{L,j})+q(\eta^\ast)}
=(c^N_{\eta^\ast})_{ij}\epsilon^{(n^{\eta^\ast}_N)_{ij}}.\nonumber
\end{eqnarray}
The Yukawa coupling $e_{ij}$ is antisymmetric: $e_{ij}=-e_{ji}$, due to presence of the antisymmetric tensor $\epsilon_{\alpha\beta\gamma}$ in the Eq. (\ref{neutrino standard+FN}).

The first line in (\ref{neutrino standard+FN}) contains the standard Yukawa interactions for the neutrinos and the two last lines  contain the additional Yukawa interactions originating from the Froggatt-Nielsen mechanism, which we will ignore due to them being suppressed.

The neutrino masses will be generated by the Yukawa terms in the first line of Eq. (\ref{neutrino standard+FN}) as the scalars obtain VEVs. The right-handed neutrino singlet $N_{R,i}$ will also have a Majorana mass term which is generated by the following operator,
\be
\mathcal{L}\supset M_0 c_{ij}^M\left(\frac{\rho^\dagger\chi}{\Lambda^2}\right)^{q(N_{R,i})+q(N_{R,j})}\overline{(N_{R,j})^{c}} N_{R,j}+h.c.,
\ee
where the mass scale $M_0$ is in principle a free parameter. We choose the mass scale to be same as the FN-messengers, in order not to introduce new mass scales into the model. 
The Majorana mass term   becomes,
\be
\mathcal{L}_\text{Majorana}=\frac{1}{2}M_{ij}\overline{(N_{R,j})^{c}} N_{R,j}+h.c.,
\ee
where the Majorana mass  matrix is,
\be\label{majorana mass}
M_{ij}=\Lambda c_{ij}^M\epsilon^{q(N_{R,i})+q(N_{R,j})}.
\ee
The messenger scale is related to the $SU(3)_L\times U(1)_X$-breaking VEVs by, 
\be
\Lambda = \sqrt{\frac{u v_2}{2\epsilon}}.
\ee
 
 The full contribution to the neutrino masses is finally given by the following terms: 
\begin{equation}
\mathcal{L}_{\textrm{neutrino mass}} = e_{ij}\epsilon_{\alpha\beta\gamma}\bar{L}_{L,i}^{\alpha}(L^{c}_{L,j})^{\beta}(\langle\eta^{\ast}\rangle)^{\gamma}+{y}^{N}_{ij}\bar{L}_{L,i}\langle\rho\rangle N_{R,j}
+{y'}_{ij}^{N}\bar{L}_{L,i}\langle\chi\rangle N_{R,j}+\frac{1}{2}M_{ij}\overline{(N_{R,j})^{c}} N_{R,j}+h.c.\label{neutrino masses}
\end{equation}
The neutrino masses can be written in a $9\times 9$ matrix form as:
\bea
\mathcal{L}_{\textrm{neutrino mass}} & = &
\frac{1}{2}\left(\overline{(\nu_L)^c}~\overline{({\nu'}_L)^c}~\overline{N_R}\right)
\left(\begin{array}{ccc}
0 & 2(m^{D})^{\dagger} &   (m^{N})^\ast\\ 
2 (m^{D})^\ast & 0 &  ({m'}^{N})^\ast\\
 (m^{N})^{\dagger} &  ({m'}^{N})^{\dagger} & M^\ast
\end{array}\right)
\left(\begin{array}{c}
\nu_L\\
\nu'_L\\
(N_R)^c
\end{array}\right)+h.c.\nonumber\\
&\equiv & \frac{1}{2}\overline{(N')^c} M_\nu
N'+h.c.,\label{neutrino mass matrix FN331}
\eea
where the $3\times 3$ sub-matrices are:
\begin{eqnarray}
m^{N}_{ij}=\frac{v_1}{\sqrt{2}} y^N_{ij},\quad {m'}^{N}_{ij}=\frac{v_2}{\sqrt{2}} y^N_{ij}+\frac{u}{\sqrt{2}} {y'}^N_{ij}
\quad \textrm{and}\quad m^{D}_{ij}=\frac{v'}{\sqrt{2}} e_{ij}\quad i=1,2,3.\label{sub-matrix definitions}
\end{eqnarray}
The mass matrix with same structure as in Eq. (\ref{neutrino mass matrix FN331}) has been studied in the literature in \cite{Catano:2012kw}.
   
We next determine the pattern of FN-charges for the leptons using the experimental values of the PMNS matrix as guidance. Once the FN-charges are known,  the exact hierarchy of the neutrino mass matrix becomes clear, and we can proceed with the block diagonalization of the neutrino mass matrix. 

\section{Neutrino masses and eigenstates}\label{Neutrino masses and eigenstates}

The Froggatt-Nielsen charges determine the hierarchy of the fermion mass  matrices. The fermion FN-charges should be chosen so that the order of magnitude of the fermion masses becomes right, thus the mass hierarchy is explained without fine-tuning. The FN charges also determine the structure of the matrices that diagonalize the fermion mass matrix. This is important as the left-handed fermion diagonalization matrices enter the two physical observables: the CKM-matrix and the PMNS-matrix. Proper  choice of  the left-handed fermion  FN-charges   can ensure that also the hierarchy of the CKM- and PMNS-matrices are produced correctly, and no fine-tuning is required. The quark sector of our model is identical to the one in \cite{Huitu:2017ukq,Huitu:2019kbm}, where  it was studied in great detail. We will therefore not consider it here. We instead concentrate on lepton sector which differs from the model presented in \cite{Huitu:2017ukq,Huitu:2019kbm} only  by the additional neutrino singlets $N_{R,i}$. 

The determining factor in our choice of leptonic FN-charges is the form of the PMNS matrix. The current experimental values of the PMNS-matrix elements by the NuFit collaboration are:
\be\label{PMNS experimental result}
\lvert U_{PMNS}\rvert=\left(\begin{array}{ccc}
0.797 - 0.840 & 0.518-0.585 & 0.143-0.156\\
0.233 - 0.495  & 0.448-0.679 & 0.639-0.783\\
0.287-0.532 & 0.486-0.706 & 0.604-0.754\\
\end{array}
\right),
\ee
where the value of  each entry is given at $3\sigma$ confidence level \cite{Esteban:2018azc}.

The PMNS-matrix elements are $\mathcal{O}(1)$ numbers in contrast to CKM-matrix where distinct hierarchy is present. The PMNS-matrix is given schematically by the left-handed charged lepton diagonalization matrix $U_L^{e}$ and the neutrino diagonalization matrix $U_\nu$ as\footnote{The exact form of the PMNS matrix is given later in Eq. (\ref{PMNS matrix without non-unitary contribution})}:
\be
U_\text{PMNS}\sim U_L^e U_\nu.
\ee
 The observed  PMNS-hierarchy is naturally obtained, if the left-handed charged lepton rotation matrix $U_L^e$ and neutrino diagonalization matrix $U_\nu$, also have this anarchical texture. This is the method  we adopt here. The hierarchy of $U_L^{e}$ and $U_\nu$ depend on the FN-charges of the left-handed leptons. The anarchical hierarchy is achieved when all the lepton families are treated equally under the FN-symmetry. We will therefore choose from now on all the lepton triplets to have equal FN-charges:
\be
 q(L^c_{L,1})=q(L^c_{L,2})=q(L^c_{L,3}) \equiv L.
\ee
 The FN-charges of the right-handed neutrino singlets do not affect the hierarchy of the light-neutrinos. We will choose the FN-charge of the right-handed neutrinos to be zero for simplicity:
\be
q(N_{R,1})=q(N_{R,2})=q(N_{R,3})=0.
\ee
 We can now see the order of magnitude in the neutrino mass matrix elements and proceed with the block diagonalization of the neutrino mass matrix.

\subsection{Neutrino mass matrices}\label{block diagonalization to light and heavier}
The neutrino mass matrix $M_\nu$ in Eq. (\ref{neutrino mass matrix FN331}) will have nine eigenvalues corresponding to nine Majorana neutrinos.
 The neutrino mass  matrix $M_\nu$  can be written in the following notation:

\begin{equation}\label{neutrino mass with blocks}
M_\nu=
\left(\begin{array}{ccc}
0 & \mathbb{M_D^T}\\
\mathbb{M_D} & \mathbb{M_R}
\end{array}\right),
\end{equation}
where
\begin{equation}\label{neutrino mass matrix blocks}
\mathbb{M_D^T}=\left(
\begin{array}{c}
2(m^{D})^{\dagger}\quad (m^{N})^\ast
\end{array}
\right)
\quad \textrm{and}\quad
\mathbb{M_R}=\left(\begin{array}{cc}
0 & ({m'}^{N})^\ast\\
({m'}^{N})^{\dagger} & M^\ast
\end{array}\right).
\end{equation}
The order of magnitude of the sub-matrices are given by
\be\label{sub-matrix estimates}
m^D_{ij}\sim v_\text{light}\epsilon^{2L+1},\quad m^N_{ij}\sim v_\text{light}\epsilon^{L+1}, \quad m^{\prime N}_{ij}\sim v_\text{heavy}\epsilon^{L} \quad  \textrm{and}\quad M_{ij}\sim v_\text{heavy},
\ee
where $v_\text{light}=v', v_1$ and $v_\text{heavy}=u, v_2$.
Note that sub-matrices in Eq. (\ref{sub-matrix estimates}) do not have an internal hierarchy, but distinct hierarchy is present between sub-matrices  $\mathbb{M_D}$ and $\mathbb{M_R}$.
The entries in the sub-matrix $\mathbb{M_R}$ are proportional to the $SU(3)_L\times U(1)_X$ breaking VEVs, $u$ and $v_2$,  whereas the entries in the sub-matrix $\mathbb{M_D}$  are proportional to $SU(2)_L\times U(1)_Y$ breaking VEVs $v'$ and $v_1$. Therefore the eigenvalues of the $\mathbb{M_R}$ are much larger than the entries in the $\mathbb{M_D}$.  This hierarchy is reflected in the eigenvalues of the matrix: it has three "\emph{light}" and six heavier eigenvalues. 

The elements in the heavier block $\mathbb{M_R}$ also have different orders of magnitude: matrix $m^{\prime N}$ is heavily suppressed compared to $M$ by  $\epsilon^L$. The eigenvalues of  the heavier block will therefore be in two distinct  scales we call "\emph{medium}" and "\emph{heavy}". Our neutrino sector is  subject to kind of "double-seesaw". The neutrino mass matrix $M_\nu$ will have in total three "light" eigenvalues, three "medium" eigenvalues and three "heavy" eigenvalues. 

The neutrino mass matrix $M_\nu$ can be block-diagonalized into three blocks, each corresponding to these eigenvalue-types according to:
\be
Z^T W^T M_\nu W Z = \left(\begin{array}{ccc}
m_\text{light}^{3\times 3} & 0_{3\times 3} & 0_{3\times 3}\\
0_{3\times 3}  & m_\text{medium}^{3\times 3} & 0_{3\times 3}\\
0_{3\times 3} & 0_{3\times 3} & m_\text{heavy}^{3\times 3}
\end{array}
\right),
\ee
where unitary matrix $W$ separates the three "light"-neutrinos from the six heavier ones, and unitary matrix $Z$  further block diagonalizes the block of heavier neutrinos into block of "medium"-mass neutrinos and "heavy" neutrinos.
The matrices $W$ and $Z$ are to the leading order: 
\be
W=\left(
\begin{array}{cc}
(1-\frac{1}{2}B_1 B_1^\dagger)_{3\times 3} & (B_1)_{3\times 6}\\
-(B_1^\dagger)_{6\times 3} & (1-\frac{1}{2}B_1^\dagger B_1)_{6\times 6}
\end{array}
\right),
\ee
and,
\be
Z=\left(
\begin{array}{ccc}
1_{3\times 3} & 0_{3\times 3} & 0_{3\times 3}\\
0_{3\times 3} & (1-\frac{1}{2}C_1 C_1^\dagger)_{3\times 3} & (C_1)_{3\times 3}\\
0_{3\times 3} & -(C_1^\dagger)_{3\times 3} & (1-\frac{1}{2}C_1^\dagger C_1)_{3\times 3}
\end{array}
\right),
\ee
with,
\bea
(B_1)_{3\times 6}&=& \Bigg(-2(m^{D})^T(({m'}^{N})^T)^{-1} M ({m'}^{N})^{-1}+(m^{N}) ({m'}^{N})^{-1}\quad  \quad 2(m^{D})^T(({m'}^{N})^T)^{-1} \Bigg)_{3\times 6}\nonumber\\
&=&\bigg( (B_1^1)_{3\times 3}\quad  (B_1^2)_{3\times 3}\bigg),
\eea
and,
\be
C_1={m'}^{N}M^{-1}.
\ee

 The light, medium and heavy blocks can be written at lowest order as:
\bea
&&m_\text{light}=
2m^{D\dagger}({m'}^{N\dagger})^{-1} M^{\ast} ({m'}^{N\ast})^{-1} 2m^{D\ast}
-[m^{N\ast}({m'}^{N\ast})^{-1}2m^{D\ast}
+2m^{D\dagger}({m'}^{N\dagger})^{-1}m^{N\dagger}]\nonumber\\
&&m_\text{medium}={m'}^{N\ast}(M^\ast)^{-1}{m'}^{N\dagger},\label{light-neutrino mass matrix in terms of blocks}\\
&&m_\text{heavy}=M^\ast.\nonumber
\eea
 
The order of magnitude of light-, medium- and heavy-neutrino masses  can now be estimated using  Eq. (\ref{light-neutrino mass matrix in terms of blocks}) with Eq. (\ref{sub-matrix estimates}):
\be\label{masses of all the neutrinos}
m_{light,ij}\sim\frac{v_\text{light}^2}{v_\text{heavy}}\epsilon^{2L+2},\quad m_{medium,ij}\sim  v_\text{heavy}\epsilon^{2L}, \quad \textrm{and}\quad m_{heavy,ij}\sim v_\text{heavy}.
\ee
 The light-neutrino masses are proportional to  $v_\text{light}^2/v_\text{heavy}$, where $v_\text{light}$ is the electroweak scale and $v_\text{heavy}$ is the scale of new physics, which is characteristic  to the  seesaw-mechanism. Additional suppression factor, $\epsilon^{2L+2}$, is however present, due to the Froggatt-Nielsen mechanism. The masses of the "medium"-neutrinos are heavily suppressed compared to $SU(3)_L\times U(1)_X$-breaking scale, making them typically lighter than $m_Z/2$. They are therefore subject to the LEP  bound \cite{Tanabashi:2018oca}-\cite{Akrawy:1990zy} on the number of light neutrinos. However, suppression on their couplings to $Z$ boson make their contribution to the invisible decay with of $Z$ boson tiny, as will be demonstrated later for our benchmark points. 
The heavy neutrinos have their masses around the  $SU(3)_L\times U(1)_X$-breaking scale and will therefore decouple.

\subsection{Neutrino eigenstates} 
The neutrino mass eigenstates are obtained once the light, medium and heavy neutrino blocks are diagonalized. The neutrino mass matrix $M_\nu$ in Eq. (\ref{neutrino mass matrix FN331}) is fully diagonalized according to:
\be\label{neutrino mass matrix full diagonalization}
M_{\nu}^{diag}=(U^T Z^T W^T) M_\nu (W Z U),
\ee
with  unitary matrix $U$ is given by,
\be
U=\left(\begin{array}{ccc}
U_{\nu}^{3\times 3} & 0_{3\times 3} & 0_{3\times 3}\\
0_{3\times  3} & U_{n}^{3\times 3} & 0_{3\times 3}\\ 
0_{3\times 3} & 0_{3\times 3}  & U_{N}^{3\times 3}
\end{array}
\right). 
\ee
The unitary matrices   $U_\nu$, $U_n$ and $U_N$  diagonalize light, medium and heavy neutrino blocks respectively.   The $U_\nu$, $U_n$ and $U_N$ are anarchical in nature,
\be\label{U texture}
U_\nu,U_n,U_N
\sim\left(\begin{array}{ccc}
1 & 1 & 1\\
1 & 1 & 1\\
1 & 1 & 1
\end{array}
\right),
\ee  
as the blocks $m_\text{light}$, $m_\text{medium}$ and $m_\text{heavy}$ have no internal  hierarchy.

According the Eq. (\ref{neutrino mass matrix full diagonalization}), the mass eigenstate neutrinos are:
\be\label{alternative gauge eigenstates}
\nu_{\textrm{mass}}\equiv 
\left(\begin{array}{c}
\nu_{light,L}^{3\times 1}\\
\nu_{medium,L}^{3\times 1}\\
\nu_{heavy,L}^{3\times 1}
\end{array}\right)
 =U^\dagger Z^\dagger W^\dagger N'.
\ee



The mixing between the neutrinos can be estimated with the use of FN-textures of the neutrino Yukawa couplings as:
\be\label{neutrino eigenstate mixing}
\left\{ \begin{array}{l}
\nu_{L}=\mathcal{O}(1)\cdot\nu_{light,L}+\mathcal{O}\left[\frac{v_\text{light}}{v_\text{heavy}}\epsilon^1\right]\cdot\nu_{medium,L}+\mathcal{O}\left[\frac{v_\text{light}}{v_\text{heavy}}\epsilon^{L+1}\right]\cdot\nu_{heavy,L},\\
\nu'_{L}=\mathcal{O}\left[\frac{v_\text{light}}{v_\text{heavy}}\epsilon^1\right]\cdot\nu_{light,L}+\mathcal{O}(1)\cdot\nu_{medium,L}+\mathcal{O}\left[\epsilon^L\right]\cdot\nu_{heavy,L},\\
(N_R)^c=\mathcal{O}\left[\frac{v_\text{light}}{v_\text{heavy}}\epsilon^{L+1}\right]\cdot\nu_{light,L}+\mathcal{O}\left[\epsilon^L\right]\cdot\nu_{medium,L}+\mathcal{O}(1)\cdot\nu_{heavy,L}.
\end{array}\right.
\ee

\section{Neutrino coupling to charged gauge bosons and PMNS-matrix}\label{Neutrino coupling to charged gauge bosons and PMNS-matrix}
Our model includes additional charged gauge boson $V_\mu^\pm$, that mixes with the $W_\mu^\pm$ boson as shown in the section \ref{charged gauge bosons}. The mixing between the charged gauge bosons is however tiny. 
The neutrino gauge eigenstates couple to the physical charged gauge bosons as:
\begin{eqnarray}
\mathcal{L}_{gCC}&=&\frac{g_{3}}{\sqrt{2}}\Big[ \bar{\nu}_{L,i}\gamma^\mu e'_{L,i}\cos\theta+\bar{\nu}'_{L,i}\gamma^\mu e'_{L,i}\sin\theta\Big]{W}^{+}_{\mu}\nonumber\\
 +&&\frac{g_{3}}{\sqrt{2}}\Big[-\bar{\nu}_{L,i}\gamma^\mu e'_{L,i}\sin\theta+\bar{\nu}'_{L,i}\gamma^\mu e'_{L,i}\cos\theta\Big]{V}^{+}_{\mu}+h.c.\label{charged gauge boson couplings}
\end{eqnarray} 
With the use of  Eqs.  (\ref{charged lepton mass matrix diagonalization convention}) and (\ref{alternative gauge eigenstates})  the coupling of  $W_\mu^\pm$ to light neutrinos can be writen as,
\bea
\mathcal{L}_{gCC}\supset\frac{g_3}{\sqrt{2}}\bar{\nu}_{\mathrm{light},L}U_\nu^\dagger\left[\left(1-\frac{1}{2}B_1 B_1^\dagger\right)\cos\theta-B_1^{1}\sin\theta\right]U_L^{e\dagger}\gamma^\mu e_L W_\mu^+ +h.c.,
\eea
from which we can identify the PMNS matrix:
\be\label{exact PMNS}
U_\text{PMNS}=\cos\theta U_L^e U_\nu -U^e_L\left[\frac{1}{2}\cos\theta B_1 B_1^\dagger+\sin\theta B_1^{1\dagger}\right]U_\nu.
\ee

The term proportional to $\cos \theta B_1B_1^\dagger$ induces nonunitarity effects to neutrino oscillations, which is an expected effect due to inclusion of sterile neutrinos in the model. Deviation from the unitarity is suppressed by the factor $\mathcal{O}(v_\text{light}^2/v_\text{heavy}^2)$ and is  significantly smaller than the current bounds \cite{Blennow:2016jkn,Escrihuela:2015wra,Antusch:2006vwa}. In any case, nonunitary mixing strength of $\geq 10^{-2}$ is ruled out.
	
The term proportional to $\sin \theta B_1^\dagger$ is similarly suppressed by a factor $\mathcal{O}(v_\text{light}^2/v_\text{heavy}^2)$, but since $B^1$ is not Hermitian, the anti-Hermitian part of it induces neutrino decay. Since the nonunitarity and unstability effects are both small, we shall ignore them in the remainder of this paper.

The PMNS matrix therefore is:
\be\label{PMNS matrix without non-unitary contribution}
U_\text{PMNS}\approx \cos\theta U_L^e U_\nu.
\ee
We have chosen the $v_\text{heavy}\gtrsim \textcolor{green}{7}$ TeV, which makes the mixing angle $\theta$ very small and $\cos\theta\approx 0.9 \approx 1$. 
As stated in Eq. (\ref{U texture}) the light-neutrino diagonalization matrix $U_\nu$ is anarchical. Since the lepton triplet FN-charges are identical also the left-handed charged lepton diagonalization matrix $U_L^{e}$ is without a hierarchy\footnote{When charged lepton mass matrix satisfies $m_{i,j}^e\leq m_{i+1,j}^e$, the left-handed diagonalization matrix satisfies: $(U_L^e)_{ij}\sim \epsilon^{\lvert q(L^c_{L,i})-q(L^c_{L,j})\rvert}$. }:
\be\label{ULe texture}
U_L^e
\sim\left(\begin{array}{ccc}
1 & 1 & 1\\
1 & 1 & 1\\
1 & 1 & 1
\end{array}
\right).
\ee  
The texture for the PMNS is therefore anarchical as well,
\be
U_\text{PMNS}
\sim\left(\begin{array}{ccc}
1 & 1 & 1\\
1 & 1 & 1\\
1 & 1 & 1
\end{array}
\right),
\ee  
which is compatible with the experimental measurements presented in Eq. (\ref{PMNS experimental result}). We note here that this is the extent which Froggatt-Nielsen setting can predict the structure of PMNS-matrix.  This is in contrast to many models involving more elaborate flavour symmetries in the neutrino sector \cite{Antusch:2007rk,King:2005bj,Masina:2005hf}. As of now, the PMNS matrix is consistent with anarchical mixing. It is up to the numerics to acquire the order-one coefficients that produce the correct lepton masses and the PMNS-matrix within the experimental limits, which is the focus of section \ref{constraints and numerical example}.

\section{Constraints and numerical examples}\label{constraints and numerical example}
There are many important experimental constraints that have to be taken into account when considering the neutrino sector of any model. Constraints for active neutrinos are the most well-known and restrictive. Least model-dependent is the direct detection bound of $m(\nu_e)$ from electron energy spectrum of tritium $\beta$ decay \cite{Kraus:2004zw} and data from supernova SN1987a burst. Also, neutrinoless double beta decay experiments \cite{Gando:2012zm}, cosmic microwave background and growth of large scale structures in the early universe \cite{Aghanim:2018eyx} all constrain the upper limits of flavour neutrino masses, and their sum. Cosmological constraints are stricter by one order of magnitude, but are dependent on the cosmological model. In addition, neutrino oscillation experiments provide neutrino mass squared differences, $\Delta m_{21}^2$ and $|\Delta m_{3j}|^2$ (with $j=1,2$ corresponding to inverted and normal mass orderings, respectively) \cite{Esteban:2018azc}. From these, a lower bound for two heavier light neutrinos can be deduced, being approximately 9 meV and 50 meV. The lightest neutrino state may be massless. Cosmological constraints are $\sum m_\nu < 0.12$~eV.

The existence of medium-mass sterile neutrinos at eV and keV scale would distort the electron energy spectrum, and different sterile neutrino mass ranges of this distortion can be detected via unstable nuclei, such as $^3$H, $^{20}$F, $^{35}$S, $^{63}$Ni and $^{187}$Re. Searches for these distortions, i.e. \textit{kink searches} have discarded large mixings of electron neutrino to medium-mass sterile neutrinos \cite{Atre:2009rg,Armbruster:2002mp}. We will show the constraints from kink searches to one of our benchmark points in Fig. \ref{electron-medium}.   The heavy TeV-scale sterile neutrinos in principle can be probed with a next-generation collider experiments, if their mixing with active flavours is large. However, our benchmark points correspond to extremely tiny TeV-scale sterile neutrino component to active neutrinos, rendering them completely inaccessible.

Our model predicts neutrinos in three different mass scales: the three sub-eV neutrinos, three heavy, mostly right-handed neutrinos, and three neutrinos \red{with masses between the sub-eV  and $SU(3)_L\times U(1)_X$ scales. The masses of the medium-scale neutrinos is determined by the $SU(3)_L\times U(1)_X$-breaking scale and in the case of $v_\text{heavy}\sim 50$ TeV the medium-scale neutrino masses are around keV scale.} The sub-eV neutrinos are constrained by their mass squared differences and mixings.  The keV neutrinos that our model predicts are constrained by the LEP bound on the number of light neutrinos, "light" here meaning neutrinos with their masses smaller than  $m_Z/2$ \cite{Tanabashi:2018oca,Abrams:1989yk,Adriani:1992zk,Akers:1994vh,Buskulic:1993ke,Adeva:1991rp,Akrawy:1990zy}.  
The coupling of  medium mass neutrinos to $Z$-bosons is heavily suppressed by the ratio between $SU(2)_L\times U(1)_Y$-  and $SU(3)_L\times U(1)_X$-breaking scales and they will pass the LEP limits on the number of light neutrinos. This becomes evident in our benchmark points. 

Seesaw mechanisms have been successfully applied to the 331 models also previously {\it e.g.} \cite{Fonseca:2016tbn,CarcamoHernandez:2017cwi}. Although it is not possible to distinguish the models experimentally from each other solely by the light neutrino sector, the particle spectra of the models may differ otherwise, for example, \cite{Fonseca:2016tbn,CarcamoHernandez:2017cwi} contain exotic charged leptons not present in the FN331 model. If the non-Standard Model particles of the model turn out to be within reach of experiments, in the case of the FN331 model, the masses and couplings are related to each other in a calculable way and thus the measurement can hint towards the model.

The tree-level CLFV decays of the charged leptons are heavily suppressed as we have stated in Section \ref{Charged lepton Yukawa couplings and masses}. The neutrinos will still mediate charged lepton decays at loop-level. These are also heavily suppressed due to small neutrino Yukawa-couplings presented in  Eq. (\ref{neutrino Yukawas in terms of epsilon}) and due to small mixing between neutrinos presented in Eq. (\ref{neutrino eigenstate mixing}). The CLFV decays of charged leptons do not pose constraints to our model, in constrast to some other 331-models such as \cite{Liu:1993gy} and \cite{Dong:2008sw}, which contain additional sources for charged lepton flavour violation such as bileptons. Also in  331-models with inverse seesaw mechanism the CLFV decays of charged leptons can be significant \cite{Boucenna:2015zwa}, \cite{CarcamoHernandez:2019vih}, in contrast to our model.

The sterile neutrinos of our model come in two distinct mass ranges "medium" and "heavy". In our benchmark points the medium sterile neutrinos are lighter than the charged leptons and cannot decay into them. The heavy sterile neutrinos on the other hand have their masses at 10-100 TeV scale and are not produced in collider experiments and therefore do not pose any bounds on our model. This is in contrast to the model in \cite{CarcamoHernandez:2019vih}.

Our model possesses the  extended  particle content of the 331-model. The additional gauge bosons and scalars of our model could potentially mediate the non-standard neutrino interactions. For example the  additional charged gauge boson, $V_\mu^\pm$, mediates the CC-NSI  given by,
\be
\mathcal{L}_{NSI}^{CC}=-2\sqrt{2}G_F \epsilon_{\alpha\beta}^{ll',L}(\bar{\nu}_{light,L,\alpha}\gamma^\mu\nu_{light,L,\beta})(\bar{l}\gamma_\mu P_L l').
\ee  
The $V_\mu^\pm$ mediated contribution to the NSI will be  heavily suppressed by its  mass:
\be
 \epsilon_{\alpha\beta}^{ll',L}\sim \frac{v_{sm}^2}{m_{V^\pm}^2},
\ee   
making it negligible as  $m_{V^\pm}\gtrsim 7$ TeV. Indeed, for our numerical benchmarks, the NSI parameters have magnitude $\mathcal{O}(10^{-13})$. All the new gauge bosons and scalars of our model have their masses proportional to the $SU(3)_L\times U(1)_X$-breaking scale. The non-standard interactions mediated by charged scalars will therefore also be suppressed due to their heavy masses.

\subsection{The FN-charges for the numerical example}
For the numerical example we take the $SU(3)_L\times U(1)_X$-breaking scale $v_\text{heavy}$ to be around \red{ $7}$ to $50$ TeV, as for  this scale the quark sector was studied in \cite{Huitu:2017ukq,Huitu:2019kbm}. We  choose the values for the leptonic FN-charges so that the  $SU(3)_L\times U(1)_X$-breaking scale is fixed.

The FN-charge of the left-handed lepton triplet $L_{L,i}$ is determined by the light-neutrino masses. According to   Eq. (\ref{masses of all the neutrinos}) all the  light-neutrino masses $m_i$  will be:
\be\label{light-neutrino masses texture estimation}
m_{i}\sim \frac{v_\text{light}^2}{v_\text{heavy}}\epsilon^{2L+2},
\ee
where the only free parameter is the FN-charge of the lepton-triplet. 

Experimentally the light-neutrino masses are constrained by \cite{Esteban:2018azc}:
\be
m_{1}<0.03\textrm{eV},\quad \Delta m_{21}^2=(7.39^{+0.21}_{-0.20})\times 10^{-5}\textrm{eV$^2$}, 
\quad \Delta m_{32}^2=(2.525^{+0.033}_{-0.032})\times 10^{-3}\textrm{eV$^2$},
\ee
where the neutrino mass squared differences are: $\Delta m_{ij}^2=m_{i}^2-m_{j}^2$.
By setting $v_\text{light}$ to electroweak scale and  $L\sim 8$ \red{or $9$}, one obtains light-neutrino masses from the correct ballpark. We will choose \red{these values} for our numerical example.

When the FN-charge of the lepton triplet is fixed, the charged lepton  mass hierarchy  depends only on the FN-charges $q(e_{R,i})$ of the right-handed charged leptons  as is evident from  Eq. (\ref{lepton-epsilon}). The FN-charges $q(e_{R,i})$ are the sole source of charged lepton mass hierarchy, as all the left-handed  lepton triplet FN-charges are identical. 
\red{We choose the right-handed charged lepton charges so that  their  mass matrix texture becomes:}
\be
m^e
\sim v'\left(\begin{array}{ccc}
\epsilon^9 & \epsilon^6 & \epsilon^4\\
\epsilon^9 & \epsilon^6 & \epsilon^4\\
\epsilon^9 & \epsilon^6 & \epsilon^4
\end{array}
\right).
\ee

 As a summary the chosen \red{lepton} FN-charges are presented in Table \ref{bps}. The FN-charges of the scalar triplets were presented in Table \ref{FN charges of scalars}.


\subsection{Numerical values for leptons}

We have chosen three benchmark points \textbf{BP1}, \textbf{BP2} and \textbf{BP3},  presented  in Table \ref{bps}. The   order-one coefficients introduced in Eq.(\ref{ce-eqs}) and Eq.(\ref{cnu-eqs}) are in the interval $\lvert c\rvert\in [0.5,5]$ to retain naturalness of the parameters. We choose different values for $SU(3)_L\times U(1)_X$-breaking VEVs $u$ and $v_2$, the $SU(2)_L\times U(1)_Y$-breaking VEVs $v'$ and $v_1$, and the FN charges for charged leptons. Below we list the explicit values for the coupling matrices we used.

\textbf{Benchmark point 1.}
\begin{align*}
c_{\eta^*}^N &=
\left( \begin{array}{ccc}
0  &  1.4094  &  4.9481\\
-1.4094   &      0  &  1.5320\\
-4.9481&   -1.5320   &      0
\end{array}\right)
&
c^N &=
\left( \begin{array}{ccc}
3.8685 &  -0.6004  &  2.5618\\
0.6590  & -0.8619   & 2.6293\\
-3.3890  & -4.1831   & 4.2469
\end{array}\right)\\
c^{'N} &=
\left( \begin{array}{ccc}
2.5272  &  3.2832  &  2.5651\\
-1.9815  &  2.3937  &  2.6579\\
-4.0777   & 4.7926   &-2.0345
\end{array}\right)
&
c^M &=
\left( \begin{array}{ccc}
4.0703&    2.3618&    1.3761\\
2.3618 &  -1.1174 &  -4.4777\\
1.3761  & -4.4777  & -2.4131
\end{array}\right)\\
c^e &=
\left( \begin{array}{ccc}
1.9737 &  -3.6284 &  -3.0577\\
-0.5178 &  -3.1689 &  -2.0976\\
-0.9467  &  3.1251  & -0.9093
\end{array}\right) &
\end{align*}

\textbf{Benchmark point 2.}
\begin{align*}
c_{\eta^*}^N &=
\left( \begin{array}{ccc}
0   & 0.5925    &4.0397\\
-0.5925  &       0 &   2.1615\\
-4.0397 &  -2.1615      &   0
\end{array}\right)
&
c^N &=
\left( \begin{array}{ccc}
3.7356&    0.5776 &   2.9485\\
0.7189 &  -1.0852  &  2.5412\\
-2.7567 &  -4.5597  &  4.1934
\end{array}\right) \\
c^{'N} &=
\left( \begin{array}{ccc}
1.9602&    2.6081 &   2.5080\\
-1.2359&    2.2310 &   2.2478\\
-3.8536 &   4.9354  & -1.6207
\end{array}\right)
&
c^M &=
\left( \begin{array}{ccc}
3.5467&    2.1954 &   1.4849\\
2.1954 &  -1.4909  & -4.3984\\
1.4849  & -4.3984   &-2.2496
\end{array}\right)\\
c^e &=
\left( \begin{array}{ccc}
2.0244&   -2.9615 &  -3.5841\\
-1.0302&   -2.3998 &  -2.0283\\
-0.7233 &   3.7798  & -1.9120
\end{array}\right) &
\end{align*}

{\textbf{Benchmark point 3.}
\begin{align*}
c_{\eta^*}^N &=
\left( \begin{array}{ccc}
0  &  -1.2219  &  4.6215\\
1.2219 &        0   & 1.3745\\
-4.6215 &  -1.3745   &      0
\end{array}\right)
&
c^N &=
\left( \begin{array}{ccc}
4.1270 & 1.2035   & 1.6042   \\
3.3340 & -1.5694  & 2.8805  \\
-2.2421 & -4.9699 & 2.6798   
\end{array}\right) \\
c^{'N} &=
\left( \begin{array}{ccc}
2.95332 & 1.54563    & 2.77676   \\
0.58451 & 3.64608    & 2.22936  \\
-3.42031 & -3.69692   & -2.68750 
\end{array}\right)
&
c^M &=
\left( \begin{array}{ccc}
3.72993  & 1.41124  & -0.66249   \\
1.41124 & -2.67661   & -3.88535  \\
 -0.66249 &  -3.88535   & -1.57821
\end{array}\right) \\
c^e &=
\left( \begin{array}{ccc}
1.5625 & -3.0524 & -3.7095  \\
-1.2578 &  -1.9733 &  -1.8557 \\
-1.5279 & 3.9513 & -1.4851  
\end{array}\right) &
\end{align*}

\begin{table}[]
	\begin{center} 
	\begin{tabular}{|c|c|c|c|}
		\hline
		\textbf{Benchmarks} & \textbf{BP1} & \textbf{BP2}              & \textbf{BP3}             \\ \hline
		$v'$ (GeV)          & 237.05       & \multirow{2}{*}{197.5999} & \multirow{2}{*}{203.205} \\ \cline{1-2}
		$v_1$ (GeV)         & 100          &                           &                          \\ \hline
                $u$ (TeV)           & 48           & 21                        & 7                        \\ \hline
		$v_2$ (TeV)         & 55           & 19                        & 7.5                      \\ \hline
$\Lambda$  (TeV )& 75.8 & 29.5 & 10.7 	\\ \hline		
		$q(L_\alpha^c)$       & \multicolumn{2}{c|}{8}                   & 9                        \\ \hline
		$q(e_R)$            & \multicolumn{2}{c|}{2}                   & 1                        \\ \hline
		$q(\mu_R)$          & \multicolumn{2}{c|}{$-1$}                & $-2$                     \\ \hline
		$q(\tau_R)$         & \multicolumn{2}{c|}{$-3$}                & $-4$                     \\ \hline
	\end{tabular}
	\caption{\label{bps}The numerical values of vacuum expectation values and FN charge assignments of leptons for our benchmarks.}
	\end{center} 
\end{table}

See Table \ref{bp-results} for the resulting neutrino masses, mass squared differences and effective strengths of nonunitary and nonstandard interactions, as well as the $V_\mu^\pm$---$W_\mu^\pm$ mixing angle.

\begin{table}[]
	\begin{center}
	\begin{tabular}{|c|c|c|c|c|}
		\hline
		\textbf{Benchmarks}                   & \textbf{BP1}                    & \textbf{BP2}                    & \textbf{BP3}                    & \textbf{Experimental values} \\ \hline
		Nonunitary strength                   & \multirow{2}{*}{$\sim 10^{-5}$} & \multirow{2}{*}{$\sim 10^{-4}$} & \multirow{2}{*}{$\sim 10^{-3}$} & \multirow{2}{*}{$ \lesssim 0.01$}   \\ \cline{1-1}
		NSI strength                          &                                 &                                 &                                 &                              \\ \hline
		$m_1$ (meV)                           & 0.0234                        & 1.85                          & 2.98  & \multirow{2}{*}{$\lesssim 55$}       \\ \cline{1-4}
		$m_2$ (meV)                           & 8.59              & 8.93                         & 9.06                           &                              \\ \hline
		$m_3$ (meV)                           & 51.2                            & 51.1                            & 50.8                            & $\lesssim 60$                        \\ \hline
		$m_1+m_2+m_3$ (meV)                    & 59.8                            & 61.9                            & 62.8  & $< 120$                      \\ \hline
		$\Delta m_{21}^2$ ($10^{-5}$ eV$^2$)  & 7.39                            & 7.64                            & 7.33                            & 6.79 --- 8.01                 \\ \hline
		$|\Delta m_{32}^2|$ ($10^{-3}$ eV$^2$) & 2.62                           & 2.61                            & 2.50                            & 2.412 --- 2.625               \\ \hline
		$m_4$ (keV)                           &  1.36                           &  0.387                    & 0.0109                                 & \multirow{6}{*}{Unknown}            \\ \cline{1-4}
		$m_5$ (keV)                           &  4.99                           &  2.03                     & 0.0206                           &                              \\ \cline{1-4}
		$m_6$ (keV)                           &  12.8                           &  5.42                     & 0.0735                            &                              \\ \cline{1-4}
		$m_7$ (TeV)                           &   184                           &  69.3                     &11.2                                 &                              \\ \cline{1-4}
		$m_8$ (TeV)                           &   380                           &  129                      &48.2                                 &                              \\ \cline{1-4}
		$m_9$ (TeV)                           &   523                           &  204                      &65.1                                 &                              \\ \hline 
		$V_\mu^\pm$ --- $W_\mu^\pm$ mixing $|\theta|$ & 0.0010 & 0.0047 & 0.014 & $\lesssim 0.01$ --- $0.04$\\ 
		\hline
	\end{tabular}
\caption{\label{bp-results}The computed values of neutrino masses, effective neutrino interaction strength and $V_\mu^\pm$---$W_\mu^\pm$ mixing for our benchmarks.}
\end{center}
\end{table}

\newpage
From the Figures \ref{electron-medium} and \ref{muon-medium} it is apparent that next-generation neutrino oscillation experiments measuring $\nu_\mu$ disappearance or neutrinoless double beta decay $(0\nu\beta\beta)$ experiments have a moderate possibility of supporting our model at \textbf{BP3}, since the present experimental limits are only approximately one degree of magnitude weaker. \green{ Of our three benchmark points, \textbf{BP3} has the greatest prospect of being detected in future, since the sterile component of $\nu_e$ has a disapprearance effect $\sum \limits_{j=4}^6 |U_{ej}|^2 \sim 10^{-3}$ and similarly the expected $\nu_\mu$ disappearance should be $\sum \limits_{j=4}^6 |U_{\mu j}|^2 \sim 10^{-4}$. For \textbf{BP1} and \textbf{BP2} the disappearance effect is smaller by a factor of $\mathcal{O}$(100) and  $\mathcal{O}$(10), respectively.}  The medium-mass neutrinos lie on the eV-scale. For the case on \textbf{BP3}, the lightest sterile neutrino $\nu'_1$, (mass $m_4$) it will be able to account the MiniBooNe anomaly \cite{Aguilar-Arevalo:2018gpe}. We calculated the active-medium neutrino mixing matrices and have illustrated them at constraint plots. Figure \ref{electron-medium} shows the constraints from $0\nu\beta\beta$ experiments \cite{Benes:2005hn,Atre:2009rg} and kink searches in single beta decay energy spectra of various unstable radioactive isotopes  \cite{Atre:2009rg} for \textbf{BP3}.  We have also included the expected sensitivity of KATRIN experiment after three-year run. Figure \ref{muon-medium} shows the constraints from muon neutrino disappearance experiments \cite{Armbruster:2002mp} and the MiniBooNe anomaly for \textbf{BP3}.

\begin{figure}[H]
	\begin{center}
		\includegraphics[width=0.65\linewidth]{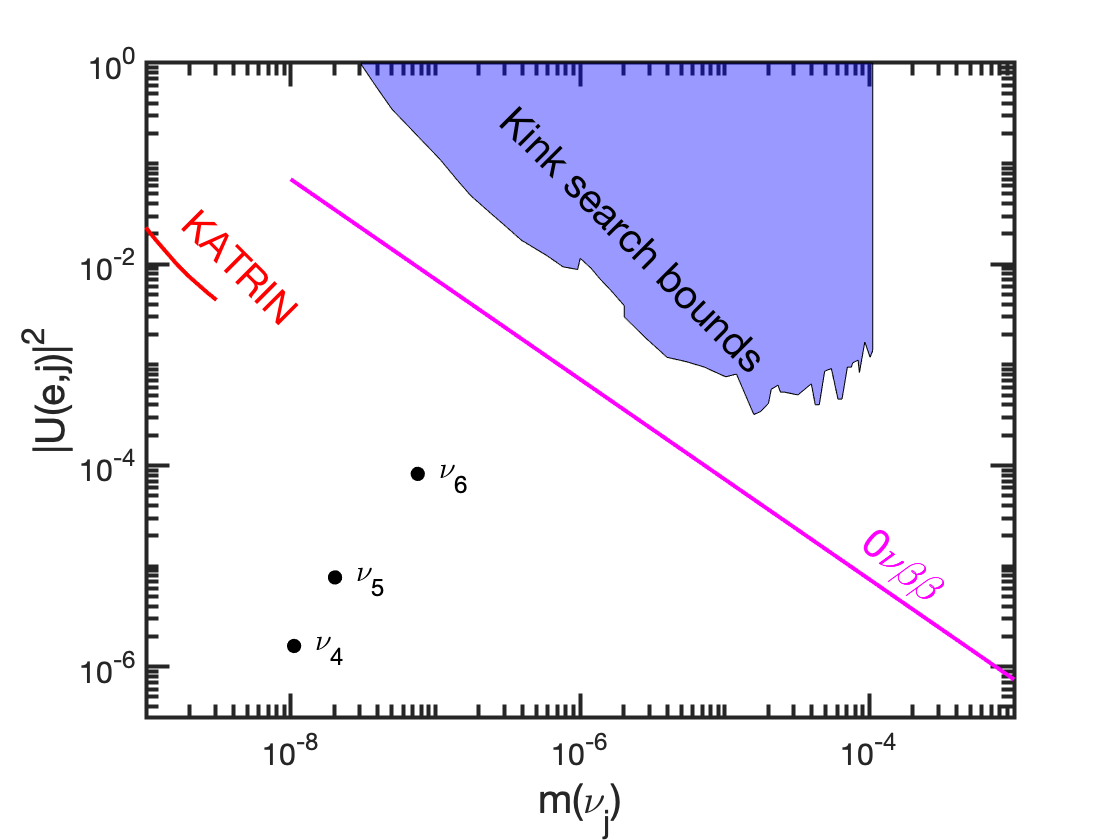}
	\end{center}
	\caption{\label{electron-medium} Constraints for the matrix element absolute values squared describing the strength of mixing of electron neutrino and medium-massive neutrinos, $U_{ej}$. Mass is in GeV units. Black dots denote the corresponding values for \textbf{BP3}.}
\end{figure}

\begin{figure}[H]
	\begin{center}
		\includegraphics[width=0.65\linewidth]{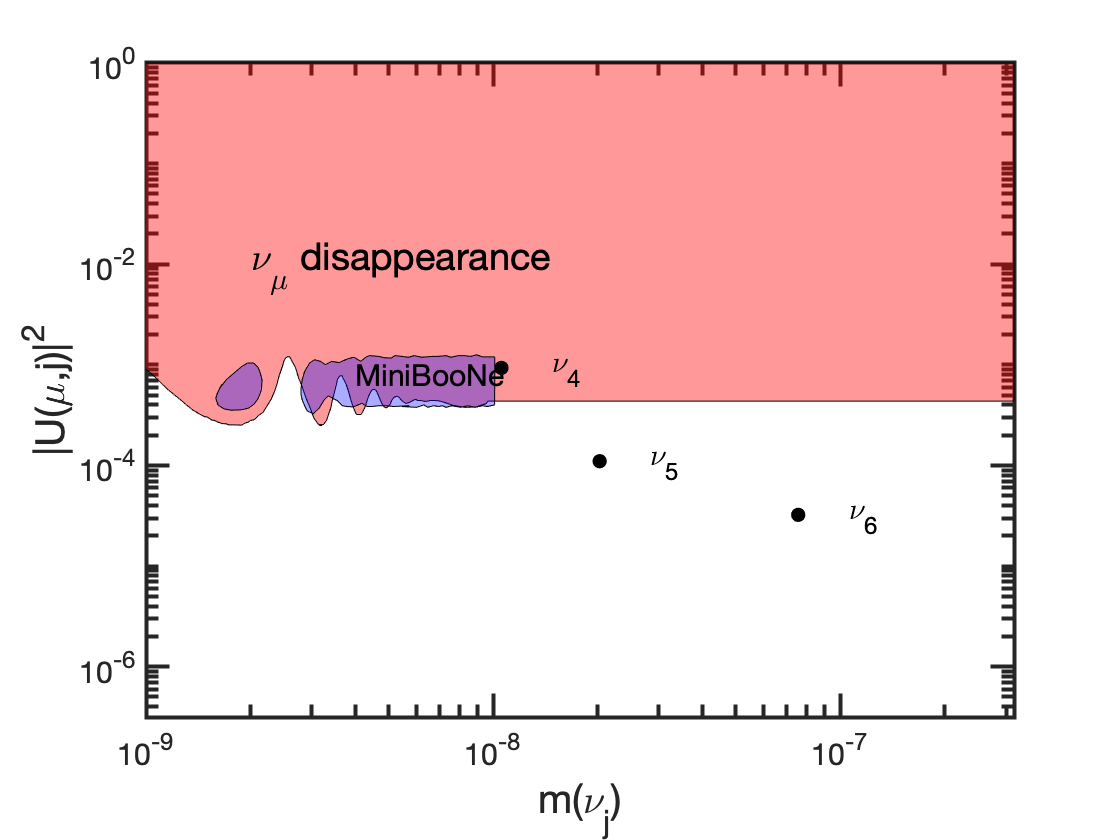}
	\end{center}
	\caption{\label{muon-medium} Same as Fig. \ref{electron-medium}, but for mixing of muon neutrinos.}
\end{figure}

\section{Conclusion}
\label{conclusion}
The FN331-model is based on $SU(3)_c\times SU(3)_L\times U(1)_X$ gauge symmetry and economically incorporates the Froggatt-Nielsen mechanism into it, thus simultaneously explaining the number of fermion families and the mass hierarchy of charged  fermions.  In this work we extended the FN331-model with three right-handed neutrino singlets. This allowed for tree-level masses for all of the neutrinos, which the original FN331-model was lacking. The neutrino masses and mixings in this model are naturally explained by  utilizing a combination of the seesaw and FN mechanisms. Lightest sterile neutrino of our model is a viable candidate for the MiniBooNe oscillation anomaly. The light-neutrino masses acquire additional suppression due to the FN mechanism, allowing the Majorana mass scale to be quite low, around few TeV. This allows for the possible collider searches of the heavy neutrinos in the future colliders. The mixing of the neutrinos, represented by the PMNS-matrix, is also explained without fine-tuning  since the FN mechanism can enforce the correct texture for the lepton mass matrices. 
As a summary the model presented here offers  an explanation for the whole fermion sector: it explains the number of fermion families and the mass hierarchy of all of the fermions, thus solving the flavour problem while fulfilling all the experimental constraints.




\vspace{0.5cm}

{\noindent \bf Acknowledgements.}  
The authors acknowledge the H2020-MSCA-RICE-2014 grant no. 645722 (NonMinimalHiggs). NK is supported by Vilho, Yrj{\"o} and Kalle V{\"a}is{\"a}l{\"a}
Foundation.

\appendix

\section{Scalar mass matrices}\label{scalar masses}

\subsection{CP-even scalars}
The CP-even scalar  mass term is
\be\nonumber
 \mathcal{L}\supset \frac{1}{2}H^T M_{cp-even}^2 H,
\ee
where  $H^T=(h_1,h_2,h_3,h_4,h_5)$ and
\begin{eqnarray}
\resizebox{1.0 \textwidth}{!} 
{
$ 
M_{cp-even}^2=
\left(\begin{array}{ccccc}
2\lambda_1{v'}^2 +f\frac{v_1 u}{v'} & \lambda_{12}v' v_1 -fu & fv_2 & \lambda_{12} v' v_2 &  \lambda_{13} v' u-f v_1\\
\lambda_{12}v' v_1 -fu   &2\lambda_2{v_1}^2 +f\frac{v'  u}{v_1} &\frac{1}{2}\widetilde{\lambda}_{23} v_2 u +b & 2\lambda_2 v_1 v_2 & \lambda_{23} v_1 u -fv'\\
fv_2 & \frac{1}{2}\widetilde{\lambda}_{23} v_2 u +b & -\frac{1}{2}\widetilde{\lambda}_{23} v_{2}^2 -b\frac{(v_1^2+v_2^2)}{v_2 u} & -b\frac{v_1}{v_2} & \frac{1}{2}\widetilde{\lambda}_{23}v_1 v_2\\
 \lambda_{12} v' v_2&2\lambda_2 v_1 v_2  &-b\frac{v_1}{v_2} &2\lambda_2 v_2^2 -b\frac{u}{v_2} &(\lambda_{23} +\widetilde{\lambda}_{23})v_2 u +b\\
\lambda_{13} v' u-f v_1&\lambda_{23} v_1 u -fv' &  \frac{1}{2}\widetilde{\lambda}_{23}v_1 v_2 &(\lambda_{23} +\widetilde{\lambda}_{23})v_2 u +b&2\lambda_3{u}^2 +f\frac{v'  v_1}{u}-b\frac{v_2}{u} 
\end{array}\right).\nonumber
$
}
\end{eqnarray}

\subsection{CP-odd scalars}
The CP-odd scalar  mass term is
\be\nonumber
 \mathcal{L}\supset \frac{1}{2}A^T M_{cp-odd}^2 A,
\ee
where  $A^T=(\xi_1,\xi_2,\xi_3,\xi_4,\xi_5)$ and
\begin{eqnarray}
M_{cp-odd}^2=
\left(\begin{array}{ccccc}
f\frac{v_1 u}{v'} & fu &0 & -fv_2 &  fv_1\\
fu   &f\frac{v'  u}{v_1} & 0 & \frac{1}{2}\widetilde{\lambda}_{23} v_2 u +b & fv'\\
0& 0&  -b\frac{u}{v_2} &  b\frac{v_1}{v_2}  & b\\
-fv_2 & \frac{1}{2}\widetilde{\lambda}_{23} v_2 u +b  &b\frac{v_1}{v_2} &  -\frac{1}{2}\widetilde{\lambda}_{23} v_{2}^2 -b\frac{(v_1^2+v_2^2)}{v_2 u} &\frac{1}{2}\lambda_{23} v_1 v_2  \\
fv_1& fv' & b &\frac{1}{2}\lambda_{23} v_1 v_2 & f\frac{v'  v_1}{u}-b\frac{v_2}{u} 
\end{array}\right).\nonumber
\end{eqnarray}

\subsection{Charged  scalars}
The charged scalar  mass term is
\be\nonumber
 \mathcal{L}\supset C^T M_{\textrm{charged scalar}}^2 C,
\ee
where  $C^T=( {\eta'}^{+},\eta^{+},\rho^{+},\chi^{+})$ and
\begin{eqnarray}
\resizebox{1.0 \textwidth}{!} 
{
$ 
M_{\textrm{ charged scalar}}^2=
\left(\begin{array}{cccc}
 f\frac{v_1 u}{v'}+\frac{1}{2}\widetilde{\lambda}_{12}v_2^2 + \frac{1}{2}\widetilde{\lambda}_{13}u^2   & \frac{1}{2}\widetilde{\lambda}_{12}v_1 v_2 & \frac{1}{2}\widetilde{\lambda}_{12}v' v_2 & \frac{1}{2}\widetilde{\lambda}_{13} v'  u +f v_1 \\
 \frac{1}{2}\widetilde{\lambda}_{12}v_1 v_2&  f\frac{v_1 u}{v'}+\frac{1}{2}\widetilde{\lambda}_{12}v_1^2  &  \frac{1}{2}\widetilde{\lambda}_{12}v'  v_1 +fu& -fv_2\\
\frac{1}{2}\widetilde{\lambda}_{12}v' v_2  &  \frac{1}{2}\widetilde{\lambda}_{12}v'  v_1 +fu & f\frac{v'  u}{v_1}+\frac{1}{2}\widetilde{\lambda}_{12}{v'}^2 &   \frac{1}{2}\widetilde{\lambda}_{23} v_2  u+b \\
 \frac{1}{2}\widetilde{\lambda}_{13} v'  u +f v_1 & -fv_2 & \frac{1}{2}\widetilde{\lambda}_{23} v_2  u +b&   f\frac{v' v_1}{u}-\frac{1}{2}\widetilde{\lambda}_{23}v_2^2 + \frac{1}{2}\widetilde{\lambda}_{13}{v'}^2-b\frac{v_2}{u}
\end{array}\right).\nonumber
$
}
\end{eqnarray}

\section{Neutral gauge boson masses}\label{neutral gauge boson masses appendix}
There are five neutral gauge bosons: $W_{3\mu}$, $W_{\mu}$, $B_\mu$, $W_{4\mu}$ and $W_{5\mu}$. 
The imaginary part of   ${X'}^{0}_\mu$   decouples from the other neutral gauge bosons and acquires a mass:
\begin{equation}
M^2_{W_5}=\frac{g_3^2}{4}(v_1^2+v_2^2+u^2).
\end{equation}

The rest of the neutral gauge bosons mix,
\be
\mathcal{L}\supset \frac{1}{2} X^T M^2_{neutral} X,
\ee
where the basis is  $X^T=(W_{3\mu}, W_{8\mu},B_{\mu},W_{4\mu})$ and
\begin{equation}
M^2_{neutral}=\frac{g_3^2}{4}\left(\begin{array}{cccc}
{v'}^2+v_1^2& \frac{v_1^2-{v'}^2}{\sqrt{3}} & -\frac{2 g_x}{3g_3}(v_1^2+2{v'}^2) &  v_1 v_2\\
\frac{v_1^2-{v'}^2}{\sqrt{3}}  &\frac{({v'}^2+v_1^2+4(v_2^2+u_2^2))}{3}&\frac{(-v_1^2+2(v_2^2+u_2^2+{v'}^2))}{ \left(\frac{3\sqrt{3}g_3}{2 g_x}\right)}  & -\frac{v_1 v_2}{\sqrt{3}}\\
 -\frac{2 g_x}{3g_3}(v_1^2+2{v'}^2) & \frac{(-v_1^2+2(v_2^2+u_2^2+{v'}^2))}{ (\frac{3\sqrt{3}g_3}{2 g_x})} & \frac{(v_1^2+v_2^2+u_2^2+4{v'}^2)}{(\frac{9g_3^2}{4g_x^{2}})} &  -\frac{4 g_x}{3g_3}(v_1 v_2)\\
 v_1 v_2& -\frac{v_1 v_2}{\sqrt{3}}  &  -\frac{4g_x }{3g_3}(v_1 v_2) & v_1^2+v_2^2+u_2^2
\end{array}\right).\nonumber
\end{equation}
The eigenvalues of this matrix can be solved analytically and they are
\begin{eqnarray}
&&m_{\gamma}^2=0,\nonumber\\
&&m_{\widetilde{W}_4}^2=\frac{g_3^2}{4}(u^2+v_2^2+v_1^2),\nonumber\\
&& m_Z ^2= \frac{g_3^2}{4}\left(\frac{3g_3^2+4g_x^2}{3g_3^2+g_x^2}\right)\left({v'}^2+\frac{v_1^2  u^2}{v_2^2+u^2}\right)+\mathcal{O}\left(\frac{v^2_{\textrm{light}}}{v^2_{\textrm{heavy}}}\right),
\nonumber\\
&& m_{Z'}^2 = \frac{3g_3^2+g_x^2}{9}(v_2^2+u^2)+\mathcal{O}\left(\frac{v^2_{\textrm{light}}}{v^2_{\textrm{heavy}}}\right).\nonumber
\end{eqnarray}
One notices  that one of the eigenvalues is exactly the same as that of the  imaginary part of the non-hermitian gauge boson. We can therefore identify the combination
\begin{equation}
X^0_\mu=\frac{1}{\sqrt{2}}(W_{4\mu}-iW_{5\mu})
\end{equation}
as the \emph{physical neutral non-hermitean gauge boson}.



\end{document}